# Design of Photonic Filters based on Integrated Coupled Sagnac Loop Reflectors

**Hamed Arianfard, Jiayang Wu, Saulius Juodkazis and David J. Moss**

*Optical Sciences Centre, Swinburne University of Technology, Hawthorn, VIC 3122, Australia.*

E-mail: *jiayangwu@swin.edu.au, dmoss@swin.edu.au*

**Abstract**

We theoretically investigate integrated photonic resonators formed by two mutually coupled Sagnac loop reflectors (MC-SLRs). Mode interference in the MC-SLR resonators is tailored to achieve versatile filter shapes with high performance, which enable flexible spectral engineering for diverse applications. By adjusting the reflectivity of the Sagnac loop reflectors (SLRs) as well as the coupling strength between different SLRs, we achieve optical analogues of Fano resonance with ultrahigh spectral slope rates, wavelength interleaving / non-blocking switching functions with significantly enhanced filtering flatness, and compact bandpass filters with improved roll-off. In our designs the requirements for practical applications are considered, together with detailed analyses of the impact of structural parameters and fabrication tolerances. These results highlight the strong potential of MC-SLR resonators as advanced multi-functional integrated photonic filters for flexible spectral engineering in optical communications systems.

# I. INTRODUCTION

Integrated photonic resonators fabricated via complementary metal-oxide-semiconductor (CMOS) technology offer competitive advantages of compact device footprint, high stability, high scalability and low-cost mass production [1, 2]. To date, these micro/nano scale resonators that provide a strong resonance field enhancement, narrowband wavelength selectivity and versatile filter shapes have found their way into many applications including for lasers, filters, modulators, switches, buffers, sensors, and signal processors [2-10].

Fano resonances, that feature asymmetric resonant lineshape profiles, are a fundamental physical phenomenon induced by interference between a discrete localized state and a continuum state [11-13]. It was first reported early in the 20th century [14, 15], and has been widely used in atom spectroscopy since then [12]. Recent advances in photonics and nanotechnology has led to new ways of realizing optical analogues of Fano resonances with broad applications in light focusing beyond the diffraction limit, optical switching, sensing, data storage, topological optics, and many others [16-20]. These optical phenomena have been demonstrated in many types of resonant cavities such as dielectric rods, disordered structures, lattices of nanospheres, metasurfaces, and integrated photonic resonators [16, 17, 19, 21-24].

Optical interleavers, switching nodes and bandpass filters (BPFs) are core components for signal multiplexing/demultiplexing, routing and monitoring in wavelength division multiplexing (WDM) optical communication systems [25-27]. To realize these devices, optical filters with a flat-top spectral response that can minimize filtering distortion, are highly desirable. To date, various schemes have been proposed to improve the roll-off of optical filters for achieving quasi flat-top spectral responses [27, 28-34]. However, these schemes, based on either finite-impulse-response (FIR) filters such as Mach–Zehnder interferometers (MZIs) or infinite-impulse-response (IIR) filters such as Fabry-Perot (FP) cavities and microring resonators (MRRs), usually achieve flat-top spectral responses via cascading many subunits [35]. This not only results in a bulky device footprint but also imposes stringent requirements on the alignment of resonant wavelengths from separate sub-components. Moreover, it is challenging to maintain the desired spectral response given the unequal wavelength drifts for different sub-components induced by the thermo-optic effect [36, 37].

To realize Fano-resonance based devices, optical interleavers as well as switching nodes and BPFs in the form of photonic integrated circuits could reap the greatest dividends in terms of compact footprint, high stability, high scalability and mass-producibility for practical applications. Recently we demonstrated multi-functional photonic filters based on cascaded Sagnac loop reflectors (CSLR) in silicon-on-insulator (SOI) nanowires [33]. Here, we theoretically investigate more advanced filter structures – namely, mutually coupled Sagnac loop reflectors (MC-SLR) − using similar principles. As compared with the CSLR resonators that include only IIR filter elements, the MC-SLR resonators that consist of both FIR and IIR filter elements provide more versatile mode interference and greatly improved flexibility for spectral engineering. We tailor the mode interference in MC-SLR resonators to achieve optical analogues of Fano resonances, or alternatively EIT or Autler-Towns splitting, that yield BPFs with ultrahigh slope rates, wavelength interleaving and non-blocking switching functions with significantly enhanced filtering flatness with improved roll-off, all in a compact footprint. Detailed analyses of the impact of varying the structural parameters, including fabrication tolerances, are provided to facilitate device design and optimization. For practical applications, the key requirements including a high extinction ratio, low insertion loss, low crosstalk and meeting the ITU-T spectral grid [38] are also considered. These results verify the effectiveness of using MC-SLR resonators as advanced multi-functional integrated photonic filters for flexible spectral engineering in optical communication systems.

## II.  DEVICE DESIGN

Figure 1(a) illustrates the schematic configuration of the MC-SLR resonators. We investigate two types MC-SLR resonators: the first, consisting of two parallel Sagnac loop reflectors (SLRs) coupled to a top bus waveguide, is termed a parallel MC-SLR resonator while the second, consisting of two inversely coupled SLRs, is termed a zig-zag MC-SLR resonator. In both resonators the bus waveguides introduce additional feedback paths for coherent optical mode interference, which provide greatly improved flexibility for engineering the spectral response. We model the MC-SLR resonators using the scattering matrix method [33, 39], where the waveguide and coupler parameters are defined in Table I. To simplify the comparison, we assume that the two SLRs are identical for each individual MC-SLR resonator, i.e., $L_{SLR1} = L_{SLR2} = L_{SLR}$, $L_1 = L_2 = L$, $t_{s1} = t_{s2} = t_s$, $t_{b1} = t_{b2} = t_b$.

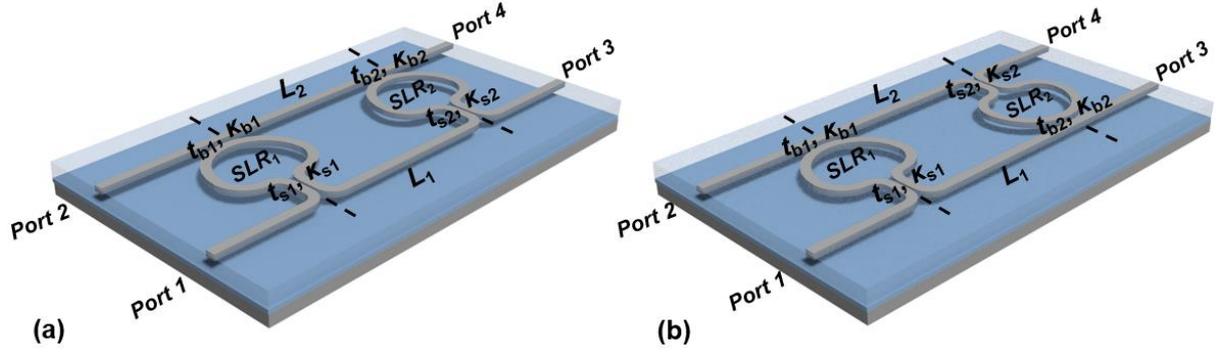

Fig. 1. Schematic configuration of (a) parallel and (b) zig-zag MC-SLR resonators consisting of two SLRs ($SLR_1$ and $SLR_2$), respectively. The definitions of $t_{si}$ ($i = 1, 2$), $t_{bi}$ ($i = 1, 2$), $L_{SLRi}$ ($i = 1, 2$), and $L_i$ ($i = 1, 2$) are given in Table I.

When ts=1, the parallel MC-SLR is equivalent to an MZI (i.e., FIR filter), and it is equivalent to a FP cavity (i.e., IIR filter) when $t_s = 1$ and $t_b = 1$, respectively. On the other hand, the zig-zag MC-SLR resonator is equivalent to a MZI combined with a SLR when $t_s = 1$ and $t_b = 1$, respectively. When $t_s \neq 1$ and $t_b \neq 1$, both can be regarded as a hybrid filter consisting of both FIR and IIR filter elements. The mutual interaction between the FIR and IIR filter elements yields a very versatile coherent optical mode interference. The freedom in designing the reflectivity of the SLRs (i.e., $t_s$), the coupling strength between the SLRs and bus waveguides (i.e., $t_b$), and the lengths of the SLRs (i.e., $L_{SLR}$) as well as the connecting bus waveguides (i.e., $L$) forms the basis for engineering the spectral response of the MC-SLR resonators, which leads to diverse applications.

TABLE I
DEFINITIONS OF STRUCTURAL PARAMETERS OF THE MC-SLR RESONATORS

| Waveguides | Length | Transmission factor [a] | Phase shift [b] |
|---|---|---|---|
| Bus waveguides between $SLR_1$ and $SLR_2$ ($i = 1, 2$) | $L_i$ | $a_i$ | $\varphi_i$ |
| Sagnac loop in $SLR_i$ ($i = 1, 2$) | $L_{SLRi}$ | $a_{si}$ | $\varphi_{si}$ |
| Directional couplers | | Field transmission coefficient [c] | Field cross-coupling coefficient [c] |
| Coupler in $SLR_i$ ($i = 1, 2$) | | $t_{si}$ | $k_{si}$ |
| Coupler between $SLR_i$ and bus waveguide ($i = 1, 2$) | | $t_{bi}$ | $k_{bi}$ |

[a] $a_i = \exp(-\alpha L_i / 2)$, $a_{si} = \exp(-\alpha L_{SLRi} / 2)$, $a$ is the power propagation loss factor.

[b] $\varphi_i = 2\pi n_g L_i / \lambda$, $\varphi_{si} = 2\pi n_g L_{SLRi} / \lambda$, $n_g$ is the group index and $\lambda$ is the wavelength.

[c] $t_{si}^2 + \kappa_{si}^2 = 1$ and $t_{bi}^2 + \kappa_{bi}^2 = 1$ for lossless coupling are assumed for all the directional couplers.

In the following sections, we tailor the spectral response of MC-SLR resonators to achieve various filtering functions with high performance, including optical analogues of Fano resonances, wavelength interleaving and non-blocking switching, and BPFs. The devices are designed based on, but not limited to, the SOI integrated platform. In our design, we use values obtained from our previously fabricated SOI devices [33, 40] for the waveguide group index of the transverse electric (TE) mode ($n_g$ = 4.3350) and the propagation loss ($\alpha$ = 55 m$^{-1}$, i.e., 2.4 dB/cm).

### III. ULTRA-SHARP FANO RESONANCES

In this section, we investigate the realization of optical analogues of Fano resonances by tailoring the mode interference in the parallel MC-SLR resonator. Figure 2(a) shows the power transmission spectrum from Port 1 to Port 4 of the parallel MC-SLR resonator. The structural parameters are $L_{SLR} = L$ = 100 μm, $t_s$ = 0.74 and $t_b$ = 0.94. It is clear that there are multiple Fano resonances with asymmetric resonant lineshapes. Figure 2(b) shows a zoom-in view of Fig. 2(a) in the wavelength range of 1550.0 nm – 1550.6 nm. The extinction ratio (ER), Q factor, and insertion loss (IL) of the Fano resonance in Fig. 2(b) are ~13.9 dB, ~33700, ~6.3 dB, respectively. In particular, the resonance spectrum shows a high slope rate (SR, defined as the ratio of the ER to the corresponding wavelength difference at the Fano resonance) of 389 dB/nm, indicating strong coherent optical mode interference in the parallel MC-SLR resonator. Compared to the FP cavity based on two cascaded SLRs [41], the top bus waveguide in the parallel MC-SLR resonator forms an additional feedback path that allows more versatile coherent optical mode interference between the two SLRs. As compared with previous work in achieving Fano resonances based on integrated MRRs [42, 43], the mutual interference between the FIR and IIR filter elements yields Fano resonances with a high SR in a compact device, requiring only two SLRs.

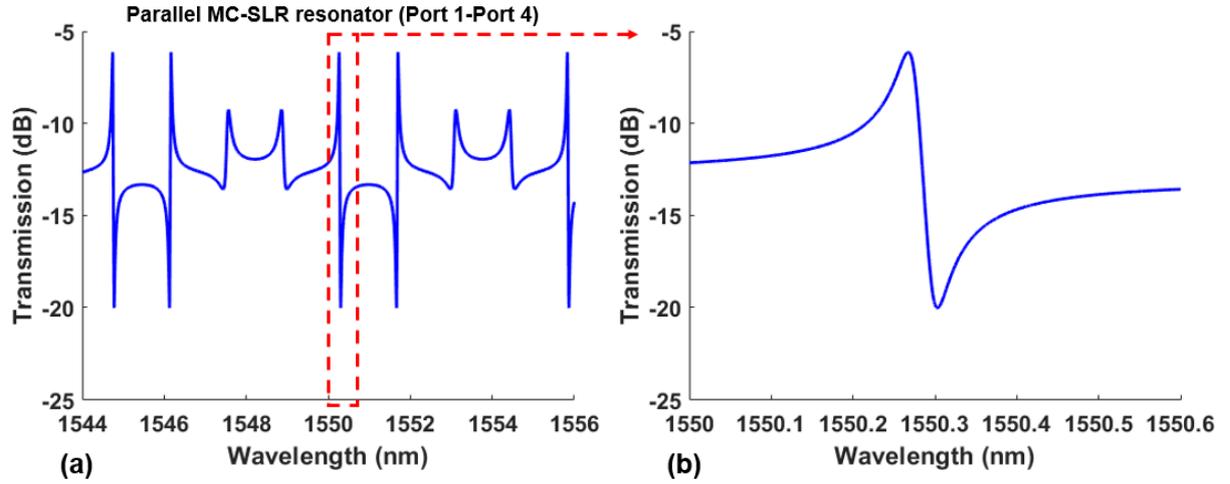

Fig. 2. (a) Power transmission spectrum of the parallel MC-SLR resonator from Port 1 to Port 4 when $L_{SLR} = L = 100$ μm, $t_s = 0.74$ and $t_b = 0.94$. (b) Zoom-in view of (a) in the wavelength range of 1550.0 nm –1550.6 nm.

In Figs. 3(a) − (c), we further investigate the impact of $t_s$, $t_b$, and $L$ on the performance of the Fano resonance generated by the parallel MC-SLR resonator. We change only one structural parameter, keeping the others the same as in Fig. 2(a). Figure 3(a-i) shows the power transmission spectra for various values of $t_s$, with the calculated SR and IL as functions of $t_s$ depicted in Fig. 3(a-ii). Clearly both the SR and the IL decrease with $t_s$, reflecting the trade-off between them. Figure 3(b-i) shows the power transmission spectra for different values of $t_b$ while the corresponding values of IL and SR are depicted in Fig. 3(b-ii). The change in SR and IL with $t_b$ shows the opposite trend to their dependence on $t_s$ while still maintaining the trade-off between them. Figure 3(c-i) shows the power transmission spectra for various $L$. Figure 3(c-ii) depicts the calculated SR and IL as functions of $L$. One can see that the resonant wavelength redshifts as $L$ increases, indicating that it can be tuned by adjusting the phase shift with thermo-optic micro-heaters [37, 39] or carrier-injection electrodes [44, 45] along the connecting bus waveguides. In Fig. 3(c-ii), both SR and IL increase with $L$, while the change in SR is more dramatic than that for IL. This indicates that the SR of the Fano resonance can be significantly improved at the expense of a slightly increased IL within a reasonable range.

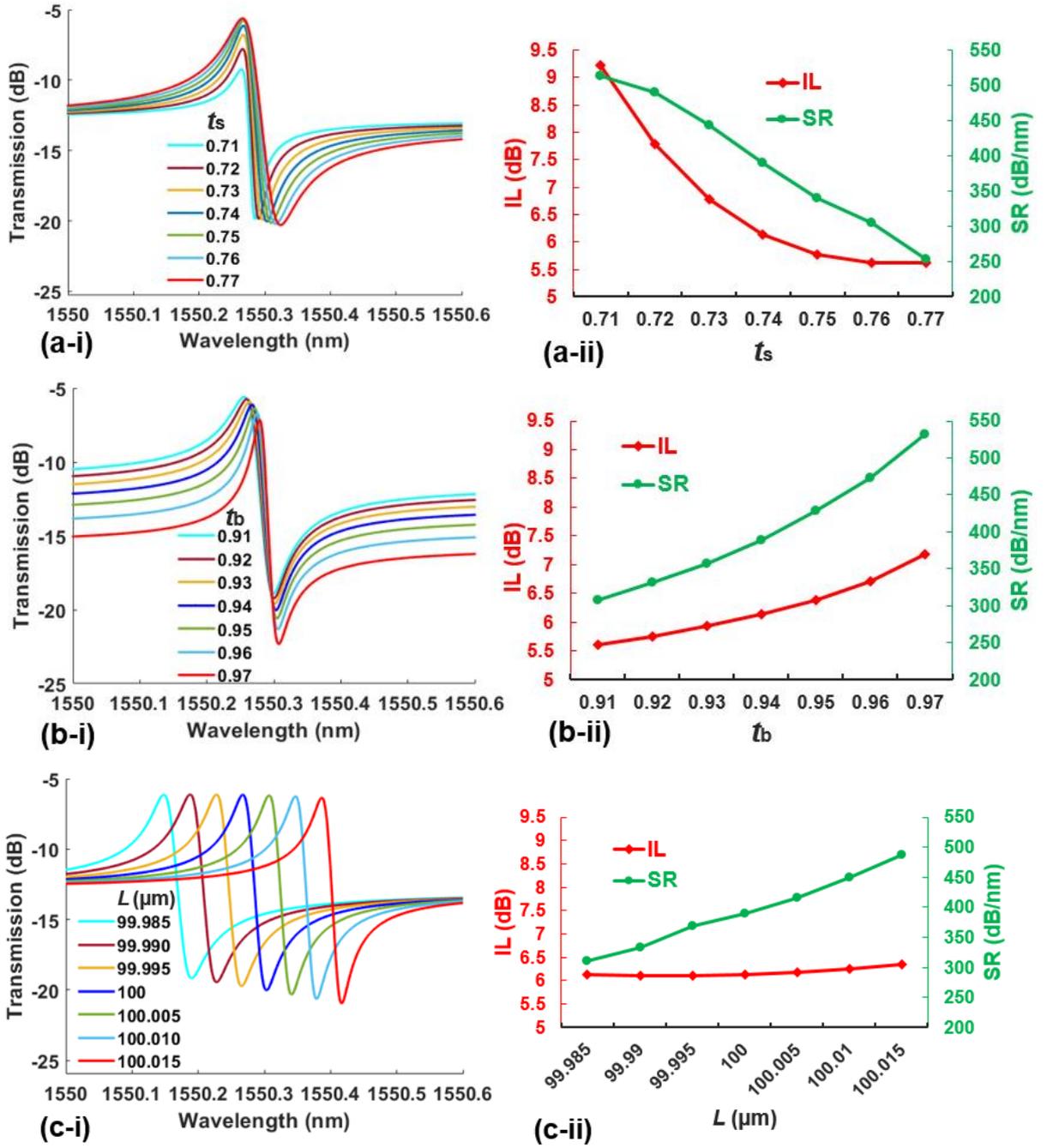

Fig. 3. (a-i) Power transmission spectra and (a-ii) the corresponding SR and IL for various $t_s$ when $t_b = 0.94$ and $L_{SLR} = L = 100$ µm, respectively. (b-i) Power transmission spectra and (b-ii) the corresponding SR and IL for various $t_b$ when $t_s = 0.74$ and $L_{SLR} = L = 100$ µm, respectively. (c-i) Power transmission spectra and (c-ii) the corresponding SR and IL for various $L$ when $t_s = 0.74$, $t_b = 0.94$ and $L_{SLR} = 100$ µm, respectively.

## IV. WAVELENGTH DE-INTERLEAVING AND SWITCHING

In this section, we investigate the use of the parallel MC-SLR resonator to achieve high performing wavelength interleaving / non-blocking switching functions in WDM optical communication systems. Figures 4(a-i) and (a-ii) show the wavelength de-interleaving operation achieved based on the parallel MC-SLR resonator with input from (i) Port 1 and (ii)

Port 2, respectively. The corresponding power transmission spectra are shown in Figs. 4(b-i) and (b-ii). The structural parameters are $L_{SLR} = L = 346$ μm, $t_s = 0.995$, and $t_b = 0.707$, which are designed in order to achieve a channel spacing (CS) of about 100 GHz in the C band to meet the ITU-T spectral grid standard G694.1 [38]. The input signal is separated into two spectrally interleaved signals that transmit to different output ports. Given the reciprocal optical transmission between the input and output ports, the same device can also perform a wavelength interleaving operation to combine the two sets of WDM signals input from Port 3 and Port 4.

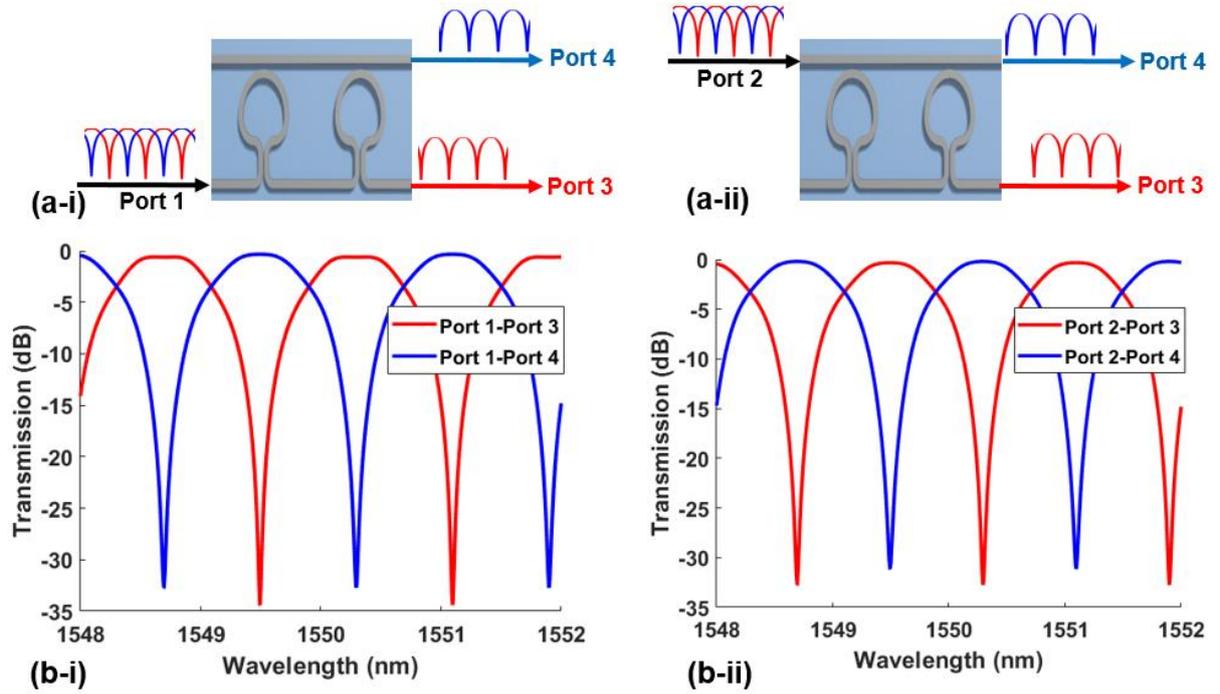

Fig. 4. (a) De-interleaving operation and (b) output transmission spectra of the parallel MC-SLR resonator, respectively. In (a) and (b), (i) and (ii) show the results for input from Port 1 and Port 2, respectively. The structural parameters are $L_{SLR} = L = 346$ μm, $t_s = 0.995$, and $t_b = 0.707$.

Figure 5(a) shows the spectral response of the parallel MC-SLR resonator for various values of $t_s$. Since the spectral responses for inputs from Port 1 and Port 2 are complementary for the same device, we only consider the spectral response from Port 1. The spectral responses at different output ports are slightly different, as shown in Fig. 5(a-i) for the output from Port 3 and Fig. 5(a-ii) for the output from Port 4. The calculated 1-dB bandwidth (BW) and normalized root-mean-square deviation (NRMSD) within the 1-dB BW range as a function of $t_s$ are depicted in Fig. 5(b). The 1-dB BW decreases with $t_s$, and the corresponding NRMSD increases with $t_s$. This reflects the deterioration of the filtering flatness for an increased $t_s$. The spectral responses of the parallel MC-SLR resonator for various $t_b$ are shown in Fig. 6(a). The 1-dB BW and the corresponding NRMSD within 1-dB BW range versus $t_b$ are plotted in Fig.

6(b). Their changes with $t_b$ shows an opposite trend to their change with $t_s$, indicating improved filtering flatness for an increased $t_b$.

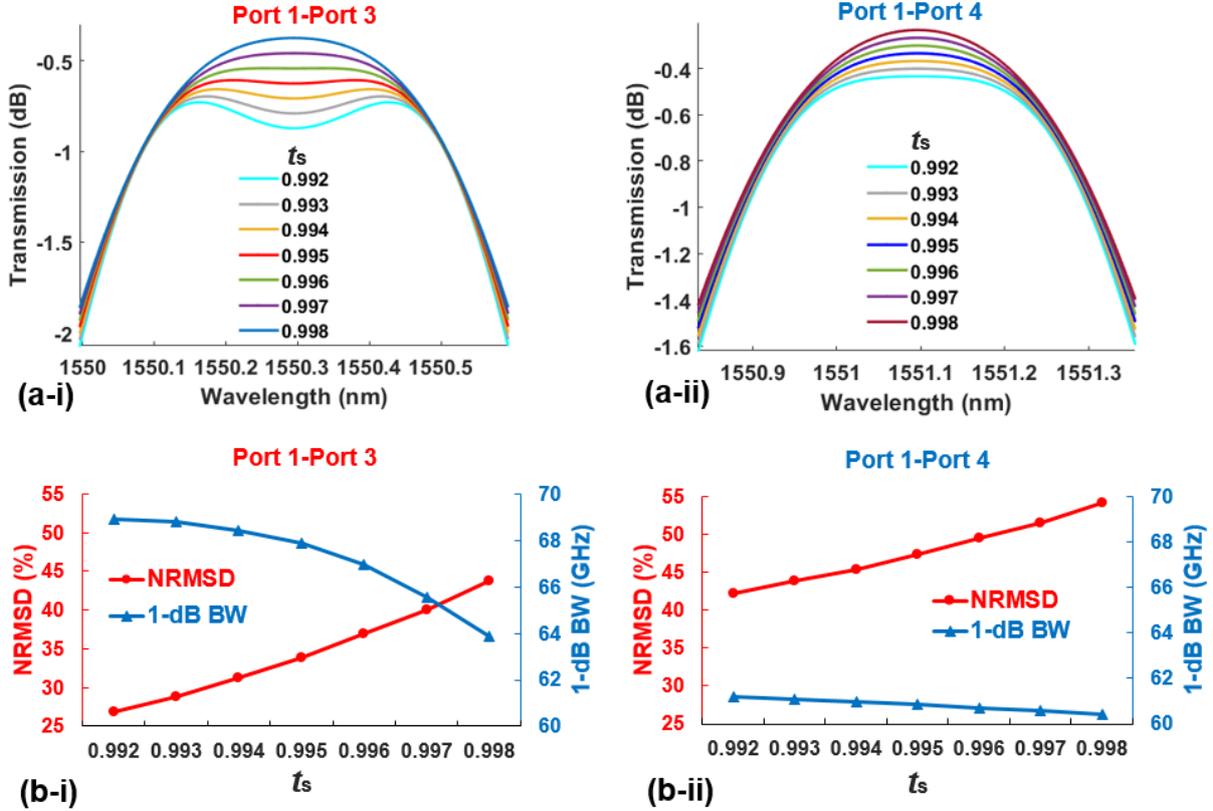

Fig. 5. (a) Power transmission spectra of the parallel MC-SLR resonator for various $t_s$ for input from Port 1 to (i) Port 3 and (ii) Port 4 when $t_b = 0.707$ and $L_{SLR} = L = 346$ µm. (b) Calculated 1-dB BW and corresponding NRMSD within 1-dB BW range as a function of $t_s$ for the transmission spectra in (a).

Table II compares the performance of the MZI and the parallel MC-SLR resonator in terms of 1-dB BW, NRMSD within 1-dB BW, ERs and ILs. The MZI and the parallel MC-SLR resonator are designed to have to a small CS of about 100 GHz. As compared with the MZI, the MC-SLR resonator shows an increased 1-dB BW and improved filtering flatness, at the expense of reduced ERs and increased ILs within reasonable ranges. Note that we used a moderately low waveguide propagation loss ($\alpha = 55$ m$^{-1}$, i.e., 2.4 dB/cm) in our design, but well within experimental capability for SOI nanowires. For waveguides with lower propagation loss, such as is achievable silicon nitride or doped silica waveguides [46-75], for example, a more significant improvement in the 1-dB BWs and filtering flatness can be achieved.

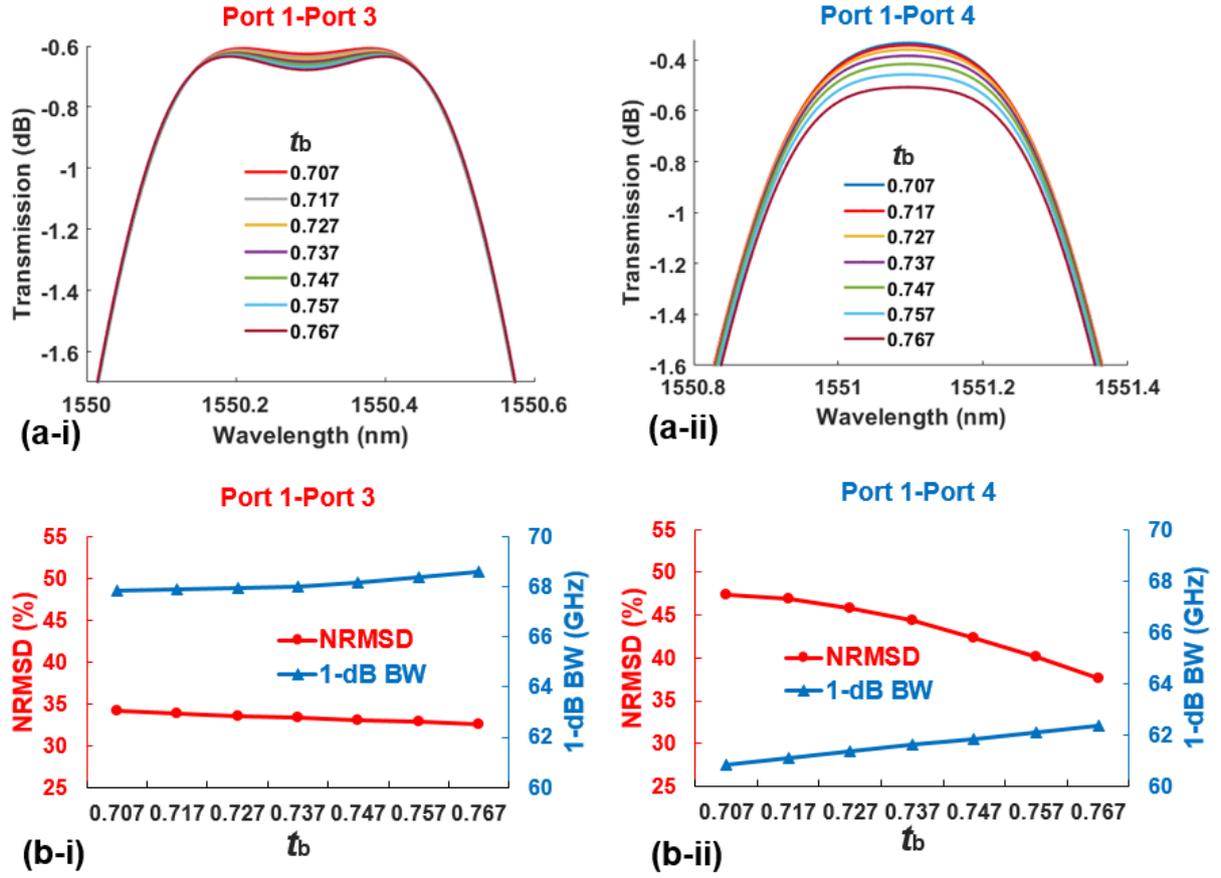

Fig. 6. (a) Power transmission spectra of the parallel MC-SLR resonator for various $t_b$ for input from Port 1 to (i) Port 3 and (ii) Port 4 when $t_b = 0.707$ and $L_{SLR} = L = 346$ µm. (b) Calculated 1-dB BW and corresponding NRMSD within 1-dB BW range as a function of $t_b$ for the transmission spectra in (a).

TABLE II
PERFORMANCE COMPARISON OF THE INTERLEAVERS BASED ON MZI AND PARALLEL MC-SLR RESONATOR

| Parameters | MZI with input light from Port 1 | | Parallel MC-SLR resonator with input light from Port 1[a] | |
|---|---|---|---|---|
| Output ports | Port 3 | Port 4 | Port 3 | Port 4 |
| 1-dB BW (GHz) | 59.9887 | 59.9767 | 67.8828 | 60.8639 |
| NRMSD within 1-dB BW (%) | 55.43 | 60.09 | 33.92 | 47.3 |
| ER (dB) | 45.9179 | 46.4525 | 33.7906 | 32.3885 |
| IL (dB) | 0.2065 | 0.1652 | 0.6083 | 0.3341 |
| CS (GHz) | 100.0053 | | 100.0053 | |

[a] The structural parameters are $L_{SLR} = L = 346$ µm, $t_s = 0.995$, and $t_b = 0.707$.

By changing $L_{SLR}$ and $L$ in the parallel MC-SLR resonator, wavelength interleaving / de-interleaving with various spectral grids can be achieved, making the MC-SLR resonator versatile enough to meet the different spectral grid requirements for different WDM systems.

Figures 7(a) and (b) show the power transmission spectra of the parallel MC-SLR resonator with CSs of approximately 50 GHz and 200 GHz, respectively. The corresponding device performance parameters and structural parameters are provided in Table III, together with those of the device with a CS of 100 GHz. The high 1-dB BW to CS ratios highlight the filtering flatness. The almost equal 3-dB BW to the CS ratios and the ERs for the complementary output ports also reflect very symmetric wavelength interleaving / de-interleaving for these devices.

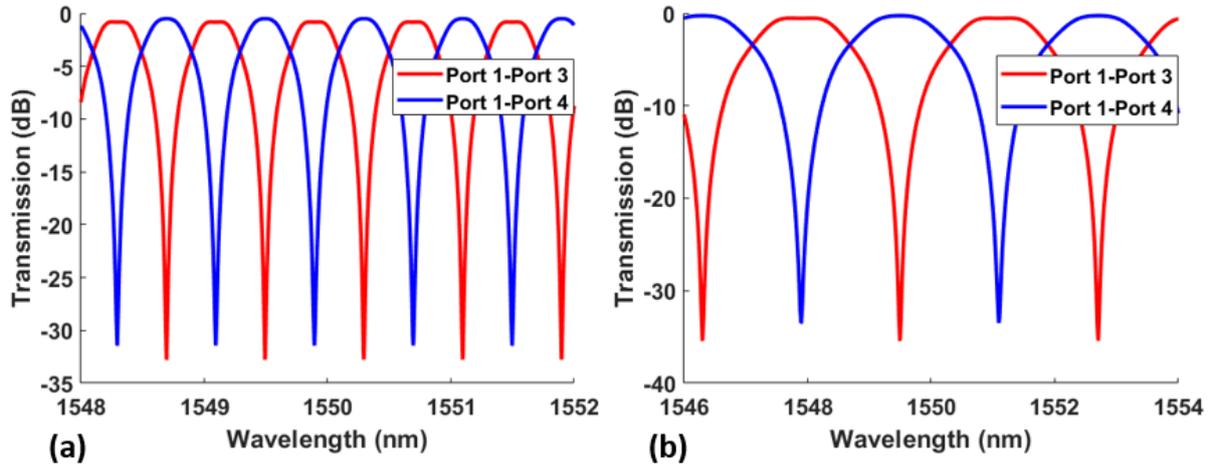

Fig. 7. Output transmission spectra of the parallel MC-SLR resonator from Port 1 to Ports 3 and 4 with CSs of (a) ~50 GHz and (b) ~200 GHz.

TABLE III
PERFORMANCE OF THE INTERLEAVER BASED ON PARALLEL MC-SLR RESONATORS WITH DIFFERENT SPECTRAL GRIDS

| CS (GHz) | | 50.0025 | | 100.0053 | | 200.0122 | |
|---|---|---|---|---|---|---|---|
| **Output ports** | | Port 3 | Port 4 | Port 3 | Port 4 | Port 3 | Port 4 |
| **ER (dB)** | | 31.9609 | 30.9357 | 33.7906 | 32.3885 | 34.8738 | 33.2153 |
| **IL (dB)** | | 0.8115 | 0.4965 | 0.6083 | 0.3341 | 0.5065 | 0.2529 |
| **1-dB BW / CS** | | 0.6773 | 0.6077 | 0.6787 | 0.6086 | 0.6780 | 0.6082 |
| **3-dB BW / CS** | | 1.0025 | 0.9923 | 1.0005 | 0.9912 | 1.0015 | 0.9917 |
| **Structural parameters** | $t_s$ | 0.995 | | 0.995 | | 0.995 | |
| | $t_b$ | 0.707 | | 0.707 | | 0.707 | |
| | $L_{SLR}$ (µm) | 692 | | 346 | | 173 | |
| | $L$ (µm) | 692 | | 346 | | 173 | |

Given the characteristics of the parallel MC-SLR resonator as a four-port device, a 2 × 2 non-blocking switching unit was further designed based on it. We chose the Benes switching architecture since it exhibits minimum complexity among various non-blocking switching architectures [76]. Figure 8(a) shows the (i) cross and (ii) bar states of the non-blocking switching unit and the corresponding spectral responses between the different ports, shown in Fig. 8(b). The structural parameters of the parallel MC-SLR resonator were the same as those in Fig. 4(b). Two resonance channels centered at wavelengths of $\lambda_1$ = 1549.4938 nm and $\lambda_2$ = 1550.2945 nm were selected for the operation of the cross and bar states, respectively. When the resonance channel at $\lambda_1$ is red shifted to $\lambda_2$, the switching unit changes from the cross state to the bar state. Practically, the red shift can be realized by slightly increasing the chip temperature via temperature controllers or injecting a high-power pump at other resonance wavelengths [26, 32, 77, 78]. Figure 8(c) shows the shift of the center wavelengths of the resonance channels at (i) $\lambda_1$ and (ii) $\lambda_2$ as a function of chip temperature variation $\Delta T$. The thermo-optic coefficient ($dn/dT$ =1.8 × 10$^{-4}$ / °C) of silicon used in our calculation was the same as that used elsewhere [76]. It can be seen that the resonance channel red shifts when increasing $\Delta T$. Table IV shows the ERs, ILs and crosstalk for the 2 × 2 non-blocking switching unit based on the parallel MC-SLR resonator. As can be seen, flat-top spectral response with high ERs, low ILs and low crosstalk is achieved. When the input port is changed to Port 2, the wavelength channels for the cross and bar states remain unchanged, i.e., $\lambda_1 = \lambda_1{'}$ and $\lambda_2 = \lambda_2{'}$.

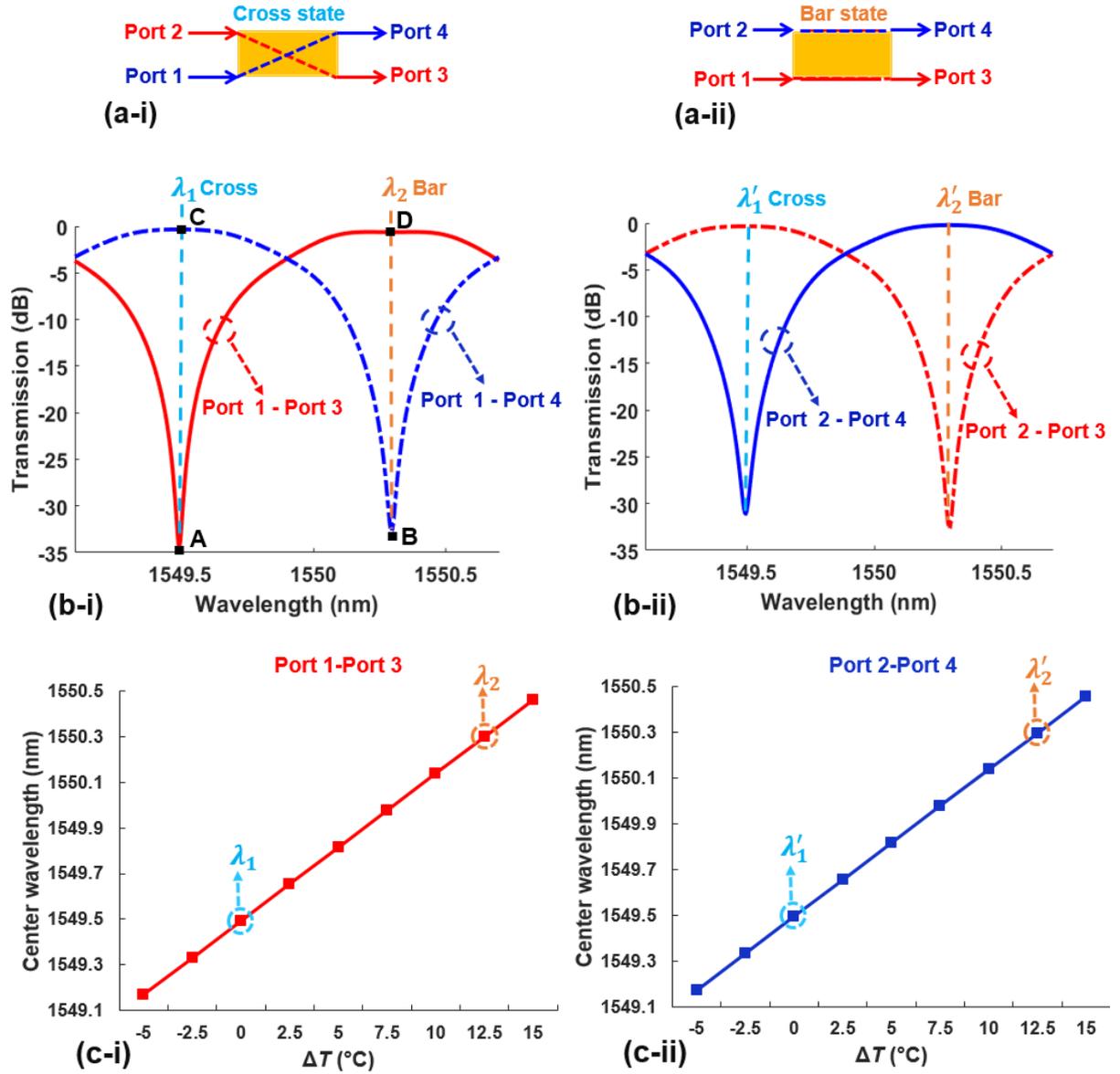

Fig. 8. (a) Schematics showing the (i) cross and (ii) bar states of a non-blocking switching unit. (b) Power transmission spectra of the parallel MC-SLR resonator (i) from Port 1 to Port 3 and Port 4 and (ii) from Port 2 to Port 3 and Port 4. $\lambda_{1,2}$ and $\lambda'_{1,2}$ denote resonance wavelengths in (b-i) and (b-ii), respectively. (c) Shift of center wavelength of the resonance channel versus chip temperature variation $\Delta T$, (i) from Port 1 to Port 3 and (ii) from Port 2 to Port 4.

TABLE IV

PERFORMANCE OF THE NON-BLOCKING SWITCHING UNIT BASED ON MC-SLR RESONATOR

| Operation state | Cross | Bar |
| --- | --- | --- |
| Operation wavelength (nm) | $\lambda_1 = \lambda_1' = 1549.4938$ | $\lambda_2 = \lambda_2' = 1550.2945$ |
| Extinction ratio (dB) | $P_C - P_B = 32.3885$ | $P_D - P_A = 33.7906$ |
| Crosstalk (dB) | $P_A - P_C = -34.0669$ | $P_B - P_D = -32.1142$ |
| Insertion loss (dB) | 0.3341 | 0.6083 |

$P_A$, $P_B$, $P_C$, and $P_D$ denote the transmission powers at point A, B, C, and D in Fig. 9b, respectively.

We also investigate the impact of varying $t_s$ and $t_b$ on the IL and ER, which are important parameters for the non-blocking switching unit. Figure 9 (a) plots the IL and ER of the transmission spectra from Port 1 to (i) Port 3 and (ii) Port 4 of the parallel MC-SLR resonator versus $t_s$. The other structural parameters were the same as those in Fig. 4(b). Clearly the IL decreases with the $t_s$ while the ER shows the opposite trend, reflecting a trade-off between them. The IL and ER as functions of $t_b$ are plotted in Fig. 9(b). As shown in Fig. 9(b), the IL of the output spectrum remains almost unchanged at Port 3 and it increase at Port 4 with the $t_b$ while the ER of the output spectrum at Port 3 decreases with the $t_b$ and it remains almost unchanged at Port 4, reflecting an increase in the difference between the ERs for different output Ports with $t_b$.

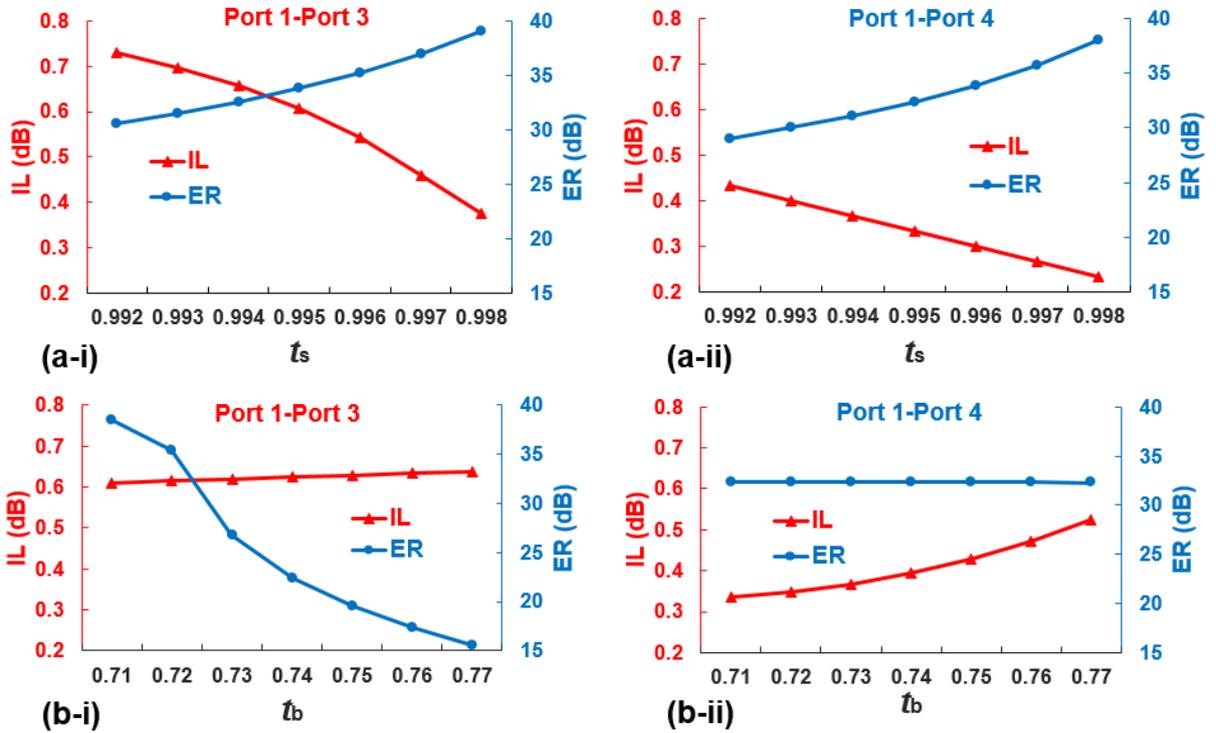

Fig. 9. (a) IL and ER of the power transmission spectra of the parallel MC-SLR resonator versus $t_s$ with input light from Port 1 when $t_b = 0.707$ and $L_{SLR} = L = 346$ µm. (b) IL and ER of output spectra versus $t_b$ with input light from Port 1 when $t_s = 0.995$ and $L_{SLR} = L = 346$ µm.

V. COMPACT BPFS WITH IMPROVED ROLL-OFF

In this section, we tailor the mode interference in the zig-zag MC-SLR resonator to realize compact BPFs with improved roll-off. Figure 10(a) shows the power transmission spectrum of the zig-zag MC-SLR resonator from Port 1 to Port 2 in the wavelength range of 1549 nm – 1550 nm. There are wide-flat stopbands and a passband with improved roll-off, arising from coherent mode interference within the zig-zag MC-SLR resonator. The structural parameters are $L_{SLR} = L = 100$ µm, $t_s = t_b = 0.78$. Figure 10(b) shows the corresponding group delay response of the BPF in Fig. 10(a). To quantitatively analyze the improvement in the filtering roll-off, we further compare the 3-dB BW of the BPF based on two zig-zag MC-SLRs (2-Z-SLRs) with BPFs based on other types of integrated photonic resonators, including a single add-drop MRR (1-MRR) [77, 78], two cascaded SLRs (2-C-SLRs) [41], three cascaded SLRs (3-C-SLRs) [33], and two parallel coupled MRRs (2-MRRs) [77, 78]. For comparison, the above filters were designed based on the same SOI wire waveguide (i.e., with the same $n_g$ = 4.3350 and α = 55 m$^{-1}$) and had the same ER and free spectral range (FSR) as those of the BPF in Fig. 10(a). Figure 10(c) shows the normalized power transmission spectra of the BPFs based on the various types of integrated resonators mentioned above. The corresponding 3-dB BWs are shown in Fig. 10(d). It is clear that the BPF based on the two zig-zag MC-SLRs resonator has the largest 3-dB BW and the best roll-off, reflecting enhanced mode interference in this compact device consisting of only two SLRs.

We further investigate the impact of $t_s$, $t_b$, and $L$ on the performance of the BPF based on the zig-zag MC-SLR resonator. We only changed one structural parameter, keeping the others the same as those in Fig. 10(a). The power transmission spectra for different $t_s$, $t_b$ and $L$ are shown in Figs. 11(a-i), (b-i), and (c-i), respectively. The corresponding ILs, ERs and 3-dB BWs are shown in Figs. 11(a-ii), (b-ii), and (c-ii), respectively. As shown in Figs. 11(a) and (b), by increasing $t_s$ and keeping constant $t_b$ or vice versa, both the ER and 3-dB BW increase, together with a slightly increased IL. This indicates that the ER and 3-dB BW can be further improved by sacrificing IL within a reasonable range. In Fig. 11(c), the ER, IL and 3-dB BW remain unchanged for different $L$ when $L_{SLR}$ is constant. This not only verifies the feasibility to implement tunable BPFs by introducing thermo-optic micro-heaters [37, 39] or carrier-injection electrodes [41, 45], but also highlights the high fabrication tolerance of the BPF. Finally, this approach towards integrated optical filters is also applicable for phase-filters such as tunable dispersion compensators and delays [79 - 83].

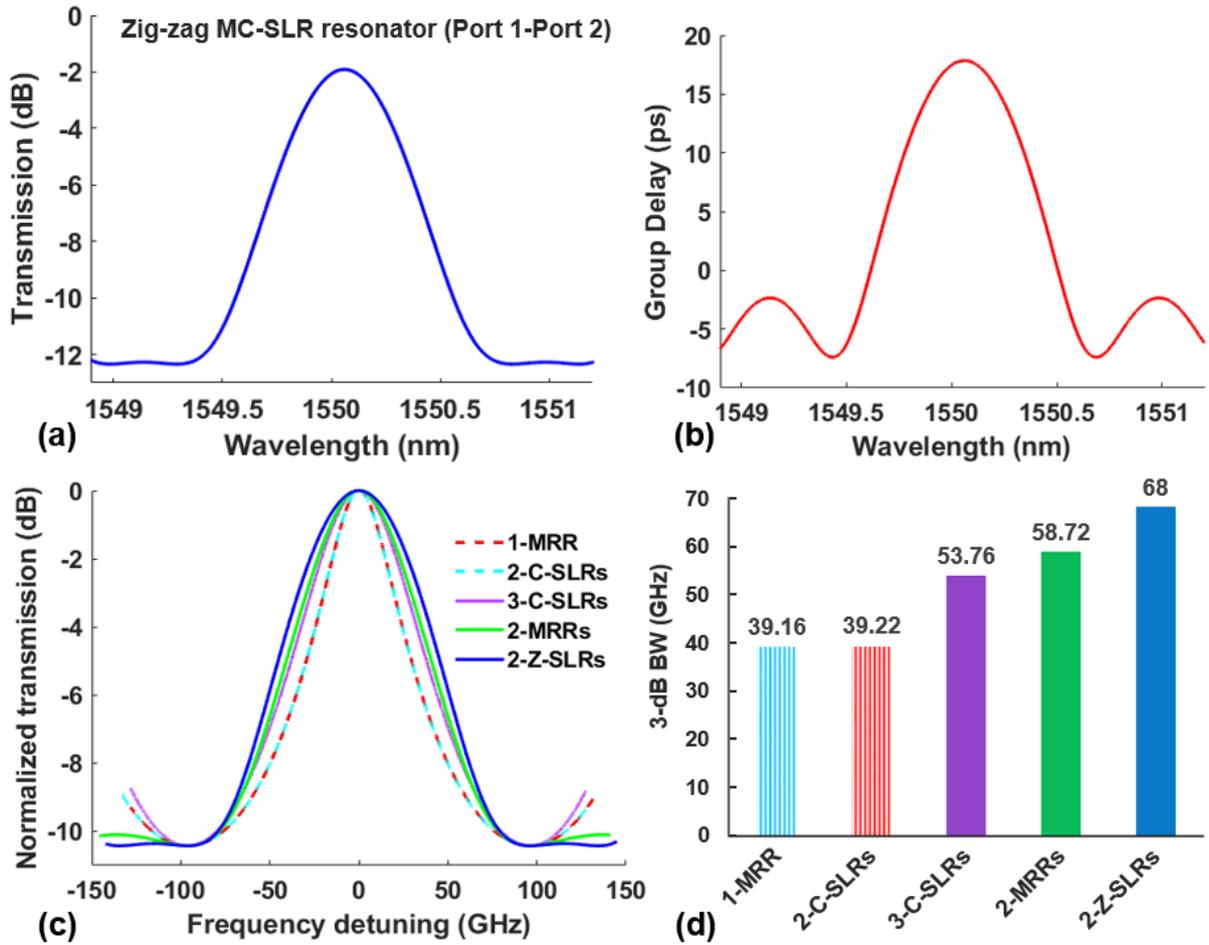

Fig. 10. (a) Power transmission spectra of the zig-zag MC-SLR resonator from Port 1 to Port 2 in the wavelength range of 1548.9 nm –1551.2 nm when $L_{SLR} = L = 100$ μm, $t_s = t_b = 0.78$. (b) Group delay of the BPF in (a). (c) Normalized transmission spectra of BPFs based on various types of integrated photonic resonators, including single add-drop MRR (1-MRR), two cascaded SLRs (2-C-SLRs), three cascaded SLRs (3-C-SLRs), two parallel coupled MRRs (2-MRRs) and two zig-zag MC-SLRs (2-Z-SLRs). (d) 3-dB BWs of the BPFs in (c).

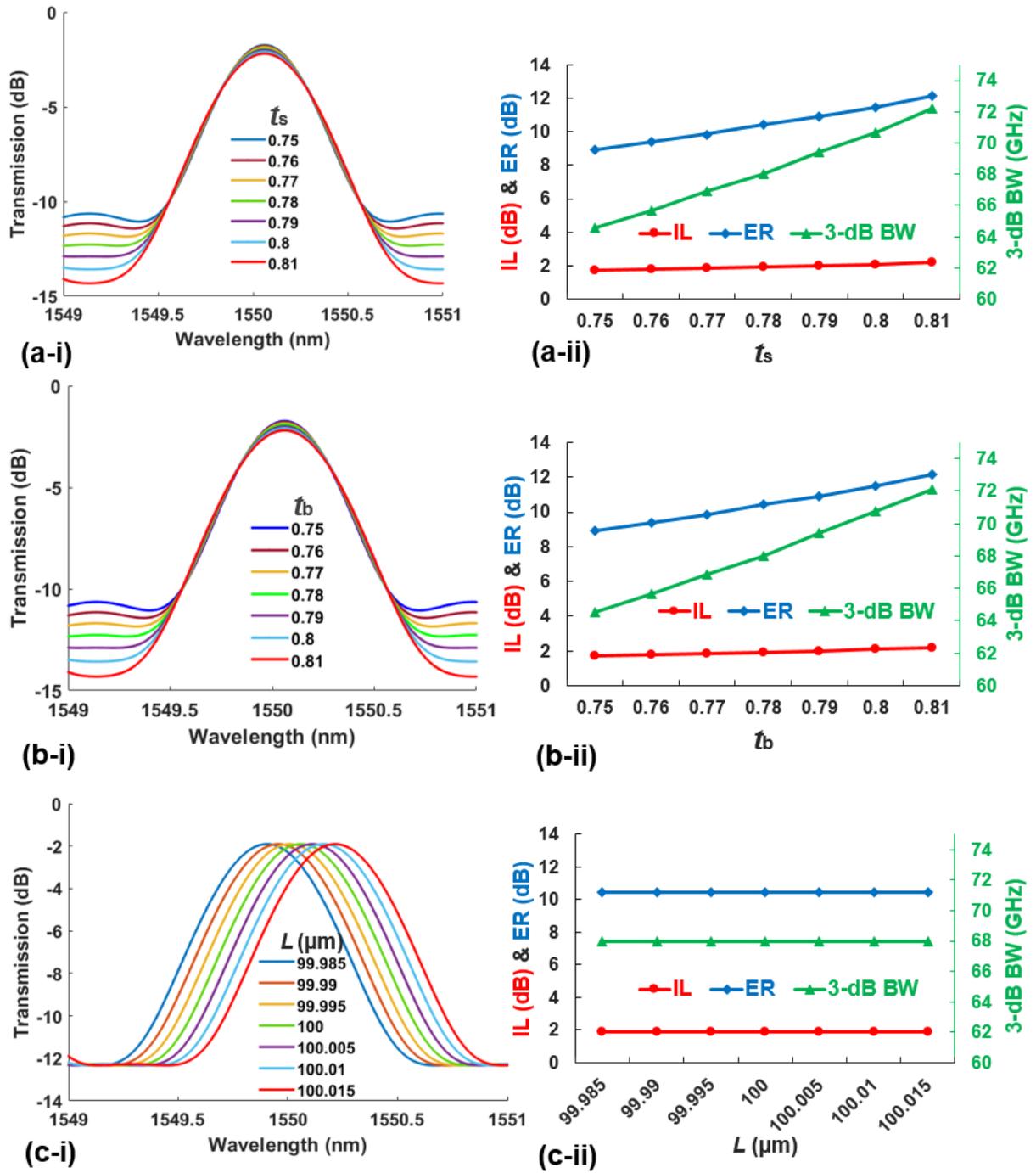

Fig. 11. (a-i) Power transmission spectra and (a-ii) the corresponding IL, ER and 3-dB BW for different $t_s$ when $t_b = 0.78$ and $L_{SLR} = L = 100$ µm, respectively. (b-i) Power transmission spectra and (b-ii) the corresponding IL, ER and 3-dB BW for different $t_b$ when $t_s = 0.78$ and $L_{SLR} = L = 100$ µm, respectively. (c-i) Power transmission spectra and (c-ii) the corresponding IL, ER and 3-dB BW for different $L$ when $L_{SLR} = 100$ µm and $t_s = t_b = 0.78$, respectively.

## VI. CONCLUSION

We theoretically investigate advanced filter structures based on MC-SLR resonators consisting of both FIR and IIR filter elements. Mode interference in the MC-SLR resonators is tailored to provide optical analogues of Fano resonances with ultrahigh SRs, a flat-top spectral response for wavelength interleaving/non-blocking switching functions, and compact bandpass filters with improved roll-off. A detailed analysis of the impact of varying the structural parameters is presented, with a particular focus on the requirements for practical applications. This work highlights the strong potential of MC-SLR resonators as multi-functional integrated photonic filters for flexible spectral engineering in optical communication systems.